Math. Inequalities and Applic. 1, N1, (1998), pp.99-104.
\input amstex

\documentstyle{amsppt}
%\NoPageNumbers
\NoRunningHeads
\magnification=1100
%\magnification=1200

\hsize5.5in
\vsize7.25in
\hfuzz=25pt
\vcorrection{.25in}

\TagsOnRight
%~
%\vskip-1in

%\centerline{PROJECT DESCRIPTION}

%\vskip.25in

\topmatter
\title Fundamental solutions to some elliptic equations with
discontinuous senior coefficients
and  an inequality for these solutions.
\endtitle

\author A.G.Ramm
\endauthor

\affil Department of Mathematics, Kansas State University,
Manhattan, KS 66506-2602, USA\\
email: {\it ramm\@math.ksu.edu}
\endaffil



\abstract
Let $Lu:=\nabla\cdot(a(x)\nabla u)=-\delta(x-y)$ in ${\Bbb R}^3$,
$0<c_1\leq a(x)\leq c_2$, $a(x)$ is a piecewise-smooth function
with the discontinuity surface $S$ which is smooth. It is proved
that in a neighborhood of $S$ the behavior of the function $u$
is given by the formula:
$$ u(x,y)=\cases
   (4\pi a_+)^{-1}[r^{-1}_{xy} + bR^{-1}], & y_3>0,\\
   (4\pi a_-)^{-1}[r^{-1}_{xy} - bR^{-1}], & y_3<0. \endcases
\tag $\ast$
$$ 
Here the local coordinate system is chosen in which the origin
lies on $S$, the plane $x_3=0$ is tangent to $S$, $a_+(a_-)$ is
the limiting value of $a(x)$ on $S$ from the half-space 
$x_3>0$, $(x_3<0)$, $r_{xy}:=|x-y|$, 
$R:=\sqrt{\rho^2+(|x_3|+|y_3|)^2}$, 
$\rho:=\sqrt{(x_1-y_1)^2 +(x_2-y_2)^2}$, 
$b:=(a_+-a_-)/(a_++a_-)$. If $S$ is the plane $x_3=0$ and 
$a(x)=a_+$ in $x_3>0$, $a(x)=a_-$ in $x_3<0$, then $(\ast)$ is
the global formula for $u$ in ${\Bbb R}^3$.
Inequality for the fundamental solution for small and large $|x-y|$ 
follows from formula $(*)$.

\endabstract

\endtopmatter

\document

\subhead 1. INTRODUCTION
\endsubhead  \vskip.2in

There are many papers on the behavior, as $x\to y$, of the
fundamental solutions to the elliptic equations of the form
$$ Lu:=\sum^n_{i,j=1} \partial_j
\left[a_{ij}(x)u_j(x,y)\right] =-\delta(x-y)
\ \text{in}\ {\Bbb R}^n,\ u_j:=\frac{\partial u}{\partial x_j}
=\partial_j u \tag 1.1  $$
for smooth coefficients $a_{ij}$. Methods of pseudo-differential
operators theory give expansion in smoothness of the solution to
(1). In [LSW] existence of the unique solution to (1) with the
properties
$$ 0<c_1r^{2-n}\leq u(x,y)\leq c_2 r^{2-n},\ 
u\in H^1_{\text{loc}}({\Bbb R}^n\backslash y),\ 
r:=|x-y|, \tag 1.2  $$
is obtained under the assumption that $a_{ij}$ are bounded 
real-valued measurable functions such that 
$$ a_1\sum^n_{i=1} t^2_i 
\leq \sum^n_{i,j=1} a_{ij}(x)t_it_j \leq a_2\sum^n_{i-1}t^2_i,
\ a_1,a_2=\text{const}>0. \tag 1.3  $$

Our purpose is to give an analytical formula for the fundamental
solution of the basic model operator (1.1), namely the operator
with 
$$ a_{ij}(x)=\delta_{ij}a(x),\qquad
a(x)=\cases 
   a_+, & x_3>0,\\
   a_-, & x_3<0. \endcases \tag 1.4   $$
Here $a_+$ and $a_-$ are positive constants, 
$$ \delta_{ij}=\cases 1, &i=j\\ 0, &i\not= j, \endcases $$
and $u(x,y)$ is the unique solution of the problem:
$$ Lu:=\sum^n_{i=1} \partial_i
\left(a(x)u_i\right) =-\delta(x-y) 
\ \text{in}\ {\Bbb R}^n\ , \tag 1.5  $$
$$[u]\vert_S=0,\quad
\left[a(x)u_N\right]\vert_S=0, \tag 1.6  $$
the symbol $[u]\vert_S$ denotes the jump of $u$ across $S$, that
is, 
$$[u]=u_+-u_-\quad, 
u_\pm:=\lim_{\varepsilon\to 0} u(s\pm \varepsilon N)\ , $$
$s$ is a
point on $S$, $N$ is the unit normal to $S$ directed along $x_3$,
$u_N$ is the normal derivative on $S$, $S$ is the plane $x_3=0$,
$[au_N]:=a_+u^+_N-a_-u^-_N$.

Problem (1.5)-(1.6) is important in many applications and is
called a transmission problem. The solution to (1.5)-(1.6) is
sought in the class $H^1_{\text{loc}}(\Bbb R^n\backslash y)\cap 
W^{1,1}_{\text{loc}}(R^n)$, where $H^1:=W^{1,2}$ and
$W^{\ell,p}_{\text{loc}}$ is the Sobolev space of 
functions whose distributional
derivatives up to the order $\ell$ belong to $L^p_{\text{loc}}$.
If 
$$a_{ij}(x)=\cases 
a^+_{ij}, & x_3>0,\\ a^-_{ij}, & x_3<0,\endcases \tag $1.4'$ $$
and the constant matrices $a^\pm_{ij}$ are positive definite,
then there exists an orthogonal coordinate transformation which
reduces $a^+_{ij}$ to $\delta_{ij}$ and
${a^-_{ij}}$ to $\lambda_j\delta_{ij}$, $\lambda_j>0$. 
We do not give the formula for $u(x,y)$ in this more general
case.

Finally note that for discontinuous coefficients equation (1.5)
is understood in the weak sense, namely as the identity:
$$ \int_{{\Bbb R}^n}a(x)u_i(x,y)\phi_i(x)\,dx=\phi(y), \tag 1.7  $$
The identity (6) for $u\in H^1_{\text{loc}}({\Bbb R}^n\backslash
y)\cap W^{1,1}_{\text{loc}}({\Bbb R}^n)$ implies conditions
(1.6).

Let $n=3$. The formula for the solution to problem (1.4)-(1.6),
or the equivalent problem (1.4), (1.7) is given in Theorem 1.1.

\proclaim{Theorem 1.1} The unique solution to problem (1.4)-(1.6)
is:
$$u(x,y)=\cases 
\frac1{4\pi a_+}\left[\frac1{r}+\frac{b}{R}\right], 
   & y_3>0,\quad b:=\frac{a_+-a_-}{a_++a_-}, \\
\frac1{4\pi a_-}\left[\frac1{r}-\frac{b}{R}\right], 
   & y_3<0,\quad r:=|x-y|,\endcases \tag 1.8  $$
$$\text{where}\quad R:=\sqrt{(x_1-y_1)^2 +(x_2-y_2)^2
+(|x_3|+|y_3|)^2}\ . $$
\endproclaim
\proclaim{Corollary 1.1} The following inequality holds
for $r\to 0$:
$$
|u(x,y)|<c|x-y|^{-1},             \tag 1.9
$$ 
where the constant $c>0$ does not depend on $x$ and $y$.
\endproclaim

Thus, the fundamental solution of the equation (1.1) with discontinuous senior coefficients has a different representation than the fundamental equation
for the similar operator with continuous coefficients, but satisfies similar
inequality for small $|x-y|$. 

A  formula, similar to (1.8) can be derived by the same method 
for $n>3$ as well. Formula (1.8)
allows one to get asymptotics of $u(x,y)$ and of $\nabla_xu(x,y)$
as $|x-y|\to 0$. Such asymptotics are useful in the study of
inverse problems for discontinuous media [3].

In section 2 we prove Theorem 1.1. In section 3 various
generalizations and applications are discussed.

\vskip.4in

\subhead 2. PROOF OF THEOREM 1
\endsubhead
\vskip.2in

The proof is given for $n=3$, but it holds with obvious small
changes for $n>3$.

The idea of the proof is to take the Fourier transform of
equation (1.5) with respect to the variables $\hat x:=(x_1,x_2)$,
to solve the resulting problem for an ordinary differential
equation analytically, and then to Fourier-invert the solution of
this problem.

Let us go through the steps.

\vskip.1in
\noindent{\bf Step 1.} 
Let $y_1=y_2=0$ without loss of generality (since $u$ is
translation-invariant in the plane 
$(x_1,x_2)$). Denote 
$$ w(\xi,x_3,y_3):=\int_{{\Bbb R}^2} e^{i\xi\cdot\hat x} 
u(\hat x,x_3; y)\,d\hat x;
\ u=\frac1{(2\pi)^2}\int_{R^2}w e^{-i\xi\cdot x}
\,d\xi. \tag 2.1  $$
$ \xi:=(\xi_1,\xi_2)$, $\xi^2=|\xi|^2=\xi^2+\xi^2_2$.
Denote $w':=\frac{\partial w}{\partial x_3}$. 

Let us Fourier-transform equation (1.5), with $a(x)$ given in
(1.4), and get
$$ w''(\xi,x_3,y_3)-\xi^2 w(\xi,x_3,y_3)=
\cases -\frac1{a_+}\delta(x_3-y_3), &x_3>0,\\
  -\frac1{a_-}\delta(x_3-y_3), &x_3<0,\endcases \tag 2.2 $$
$$w(\xi,+0,y_3)-w(\xi,-0,y_3)=0,\quad  
a_+w'(\xi,+0,y_3)-a_-w'(\xi,-0,y_3)=0. \tag 2.3 $$

In what follows we omit $\xi$ in the variables of $w$, and write
$w(x_3,y_3)$ for brevity. Thus, $w$ solves problem (2.2)-(2.3)
and satisfies the condition
$$w(\pm\infty,y_3)=0\ . \tag 2.4  $$

Assume that $y_3\not=0$. Then problem (2.2)-(2.4) has a solution
and this solution is unique. A lengthy but straightforward
calculation yields the formula for $w$:
$$ w=\cases
\frac{\exp\left(-|\xi|x_3-y_3|\right)}{2|\xi|a_+} 
 +b\frac{\exp\left[-|\xi|\left(|x_3|+|y_3|\right)\right]}
   {2|\xi|a_+},   &y_3>0\\
\frac{\exp\left(-|\xi||x_3-y_3|\right)}{2|\xi|a_-}
 -b\frac{\exp\left[-|\xi|\left(|x_3|+|y_3|\right)
\right]}{2|\xi|a_-}, &y_3<0 \endcases  \tag 2.5  $$
$$\text{where}\quad b:=\frac{a_+-a_-}{a_++a_-},\qquad a_-,a_+>0.
  \tag 2.6  $$

\vskip.1in
\noindent{\bf Step 2.} 
The function $u(x,y)$ is obtained from $w$ by the second formula
(2.1). Let us denote $|\xi|:=v$, 
$\rho:=|\hat x|=\sqrt{x^2_1+x^2_2}$, and
remember that $y_1=y_2=0$. Since $w$ depends on $|\xi|$ and does
not depend on the angular variable, $|\xi|:=v$, we have
$$ u=\frac1{(2\pi)^2}\int^\infty_0\,d\nu\, \nu\int^{2\pi}_0 
e^{-i\nu\rho\cos\varphi}w\,d\varphi =\frac1{2\pi} \int^\infty_0\,
ds\, \nu wJ_0(\nu\rho) \tag 2.7 $$
where $J_0(x)$ is the Bessel function and we have used the known
formula:
$$\frac1{2\pi}\int^{2\pi}_0
e^{iv\rho\cos\varphi}d\varphi=J_0(v\rho). $$
We need another well-known formula:
$$ \int^\infty_0 e^{-\nu t}
J_0(\nu\rho)d\nu=\frac1{\sqrt{\rho^2+t^2}},\qquad t>0 \tag 2.8 $$
From (2.5), (2.7) and (2.8) we get (1.8) with $y_1=y_2=0$.
Therefore, recalling the translation invariance of $u$ in the
horizontal directions, we get (1.8).

Theorem 1.1 is proved. \qed

\noindent{\bf Remark 2.1}
Note that the limits of $u(x,y)$ as $y_3\to\pm 0$ exist and are
equal:
$$ u(x,\hat y,+0)=u(x,\hat y,-0)= 
\frac1{2\pi r(a_++a_-)}. \tag 2.9 $$
A result similar to (2.9) is mentioned in [K, p.318], however the
argument [K] is not clear: the differentiation is done in the
classical sense but the functions involved have no classical
derivatives: they have a jump.

\vskip.4in

\subhead 3. GENERALIZATIONS, APPLICATIONS
\endsubhead
\vskip.2in

This section contains some remarks. 

\vskip.1in

\noindent{\bf Remark 3.1}
First, note that if $a(x)$ is a piecewise-smooth function with a
smooth discontinuity surface $S$, $s\in S$,
$a_\pm=\lim_{\varepsilon\to 0}a(s\pm\varepsilon N)$, where $N$ is
the exterior normal to $S$ at the point $s$, then the main term
of the asymptotics of the fundamental solution $u(x,y)$ in a
neighborhood $U_s$ of the point $s\in S$ is given by formula
(1.8) in which $x,y\in U_s$. This follows from the fact that the
main term in smoothness of the solution to an elliptic equation
in $U_s$ is the same as to the equation with constant
coefficients which are limits of $a(x)$ as $x\to s$. In our case,
this ``frozen-coefficients" model problem is given by equations
(1.4)-(1.6). This argument shows that the same conclusion holds
if the coefficient $a(x)$ in ${\Bbb R}^3_+$ and in 
${\Bbb R}^3_-$ is not smooth but just Lipschitz-continuous.

\noindent{\bf Remark 3.2}
In principle, our method for calculation of $u(x,y)$ for the
model problem (1.4)-(1.6) is applicable for the model problem
$(1.4')$ with anisotropic matrix.

\noindent{\bf Remark 3.3}
We only mention that our result concerning asymptotics of
$u(x,y)$ as $|x-y|\to 0$ for piecewise-smooth coefficients is
applicable to inverse problems of geophysics and inverse
scattering problems for acoustic and electromagnetic scattering
by layered bodies.

For example, if the governing equation is [R, p.14]:
$$\nabla\cdot[a(x)\nabla u]+k^2q(x)u=-\delta(x-y)\quad \text { in  }
{\Bbb R}^3, $$
$k=\text{const}>0$, say $k=1$, $q(x)=1+p(x)$, where $p(x)$ is a
compactly supported real-valued function, 
$p(x)\in L^2_{\text{loc}}({\Bbb R}^3)$, 
$\text{supp} (p(x))\subset {\Bbb R}^3_-:=\{x:x_3<0\}$,
$a(x)=1+A(x)$, where $A(x)$ is compactly supported 
piecewise-smooth function with finitely many closed compact
smooth surfaces $S_j\subset {\Bbb R}^3_-$ of discontinuity.
Across these surfaces the transmission conditions (1.6) hold, and
at infinity $u$ satisfies the radiation condition. Then $u(x,y)$
is uniquely determined. 

An inverse problem is: given $g(x)$ and
$u(x,y)$ for all $x,y\in S:=\{x:x_3=0\}$ and a fixed $k=1$, can
one uniquely determine $a(x)$, in particular, the discontinuity
surfaces $S_j$?

To explain how Theorem 1.1 can be used in this
inverse problem, note that if two systems of surfaces $S^{(1)}_j$
and $S^{(2)}_j$ and two functions $a_1$ and $a_2$ produce the
same surface data on $S$ for all $x,y\in S$, then an
orthogonality relation [R, pp. 65, 86] holds:
$$\int v(x)\nabla u_1(x,y)\nabla u_2(x,z)\,dx=0,\qquad 
\forall y,z\in D'_{12}, \tag 3.1 $$
where $v(x)=a_1-a_2$, $u_m(x,y)$, $m=1,2$, are the fundamental
solutions corresponding to the obstacle $D_m$ (i.e., to $a_m$ and
$S^{(m)}_j$), $D'_{12}:={\Bbb R}^3\backslash D_{12}$,
$D_{12}:=D_1\cup D_2$. 

Let us prove, e.g., that 
$\partial D_1=\partial D_2$, using (3.1). If there is a part of
$\partial D_1$ which lies outside $D_2$, and $s$ is a point at
this part, then, assuming (for simplicity only) that $v(x)$ is
piecewise-constant and using for $\nabla u_1$ and $\nabla u_2$
formulas, which follow from (1.8) as $y=z\to s$, we conclude that
the left-hand side of (3.1) is an integral which contains a part,
unbounded as $y=z\to s\in\partial D_1$: $c\int|x-y|^{-4}dx$,
$c=\text{const}\not= 0$. This contradicts to (3.1). Therefore
there is no part of $\partial D_1$ which lies outside $D_2$.
Likewise, there is not part of $\partial D_2$ which lies outside
$\partial D_1$. Thus, $\partial D_1=\partial D_2$. Similarly one
proves that $S^{(1)}_j=S^{(2)}_j$ for all $j$, provided (3.1)
holds.

A detailed presentation of such an argument is given in the paper 
by C. Athanasiadis, A.G. Ramm and I. Stratis, Inverse acoustic
scattering by layered obstacle (in preparation).

%\vskip.5in

\vfill

%\noindent email: ramm\@math.ksu.edu

\pagebreak

\enddocument